\documentstyle[epsf,twocolumn,floats,prl,aps]{revtex}    
\newcommand{\figwidth}{3.375in}     
\begin{document}
\draft

\twocolumn[\hsize\textwidth\columnwidth\hsize\csname @twocolumnfalse\endcsname

\title{Phase diagram of second layer of $^4$He adsorbed on graphite 
}
\author{ Marlon Pierce and Efstratios Manousakis}
\address{
Department of Physics and Center for Materials Research and Technology,
Florida State University, Tallahassee, FL 32306-4350
}
\date{\today}
\maketitle
\begin{abstract}
\noindent
Using realistic helium-helium and helium-graphite interactions and
the path integral Monte Carlo method, we are able to
identify gas, superfluid liquid, 
commensurate-solid, and incommensurate- 
solid phases, and the coexistence regions between them,
for the second layer of $^4$He on graphite.
The phase boundaries and the specific heat 
are in good agreement with experiment.
The appearance and disappearance 
of superfluidity with increasing coverage can be explained by 
the growth of coexistence phases, as was observed by 
torsional oscillator experiments.
\end{abstract}

]


Films of $^4$He adsorbed on graphite have a very rich phase diagram and
provide an excellent realization
of nearly two-dimensional (2D) phenomena.  Several interesting 
phases occur, including
fluid phases, a variety of commensurate structures, and 
incommensurate-solid phases\cite{dash78,schick80,greywall}.  
These phases and the coexistence regions that separate
them are governed by a delicate balance of adatom and substrate interactions.
Furthermore, the large zero-point motion of the
helium atoms implies that quantum 
effects such as particle permutations play an important role in the 
phase diagram.  

Many experimental studies of the helium-graphite system have been
performed. Heat capacity
measurements\cite{dash78,schick80,greywall,bretz78} show that at low
temperatures the first- and second-layer phase diagrams are similar,
progressing with increasing density through gas, liquid,
commensurate-solid, and incommensurate-solid phases, with
coexistence regions separating these uniform phases.  Neutron
scattering\cite{nielsen80,lauter,carneiro81} can detect the
commensurate first-layer solid and the incommensurate first- and
second-layer solids, but no direct evidence for the structure of the
second-layer commensurate solid exists.  It is believed\cite{greywall}
to be in $\sqrt 7 \times \sqrt 7$ partial registry with the
first-layer helium solid, in analogy with $^3$He on
graphite\cite{elser89}.  These experiments are supplemented by
torsional oscillator (TO) measurements \cite{reppy}, which
detect superfluidity only in the second and higher layers.
The second layer thus presents a unique opportunity to study the
interplay of superfluid and solid phases in two dimensions.  

Superfluidity is caused by particle-permutation cycles of infinite 
length.  Permutations
apparently do not play an important role in the
first layer because no superfluidity has been detected, but are 
very important in the second and 
higher layers, which do have superfluid phases.  
Without including particle permutations, which simplifies
the simulation dramatically, the Monte Carlo simulation of
Ref. \cite{abraham87}
reproduced most of the interesting first-layer features.  
This provides additional
evidence that permutations are not important in the first layer.
In the second layer, the
commensurate-solid phase of $^3$He 
has also been simulated without 
permutations\cite{abraham90}, but one needs to simulate particle permutations
in addition to particle moves in order to allow for the possibility that
a superfluid phase may be found.
In addition, it is expected
that the other second-layer phases and their boundaries
will be effected by the inclusion of particle permutations.

Using realistic helium-helium\cite{aziz92} and helium-graphite\cite{cole80}
interactions and a path integral Monte Carlo (PIMC) method
for simulating strongly correlated Bose systems that
includes particle permutations,
we have examined
the second layer of $^4$He on 
graphite.  For the first time with
simulation, we are able to identify coverage regions 
where this system is in gas (G), 
superfluid liquid (L), commensurate-solid
(C), and incommensurate-solid (IC) phases, and the coverage regions
of the coexistence phases that separate them, namely, the G-L, L-C and C-IC
phases.  The realistic treatment of the substrate and first layer 
is needed to produce
the C phase, which is absent in
2D calculations\cite{whitlock88}.
The phase boundaries are in reasonable agreement with
heat capacity and torsional oscillator measurements
\cite{greywall,reppy}.  
The experimentally observed reentrant superfluidity can be explained by this
phase diagram.
Superfluidity appears as increasing coverage causes a transition from
gas-liquid to liquid and disappears at still higher coverage with the
growth of liquid-commensurate solid coexistence.
We further present the first simulation results for the 
superfluid phase and the first direct evidence for 
the $\sqrt 7 \times \sqrt 7$ solid for $^4$He on graphite.
Finally, we obtain the specific heat for the
L, C, and IC phases
and find peaks at temperature values that
are in reasonable agreement with experiment.

In the PIMC method 
both the spatial configurations of 
the particles and the possible permutations of particle labels
must be sampled.  
A detailed outline of the application of PIMC to $^4$He systems can
be found in Refs. \cite{ceprev}.  
We have 
developed a PIMC method based on these references
and have 
tested it on bulk helium, reproducing the 
energy, specific heat, and superfluid density given in 
Refs. \cite{cep86,runge92}. Below, we briefly summarize how we have 
extended the method for simulating layered systems on a substrate.
A detailed description  
will be given in a forthcoming publication\cite{forthcom}. 
An alternative approach for applying PIMC to films
can be found in Refs. \cite{wagner92,wagner96}.
We model the graphite substrate 
as a featureless slab, so the
effective helium-graphite interaction depends only on the height of the helium
above the substrate\cite{cole80}.  On the substrate we first place a layer of
helium atoms at fixed height, frozen
at triangular lattice sites.
The first-layer height is set at the graphite's potential 
minimum, 2.8 $\AA$, and the 
density is fixed at its compressed 
value, 0.127 atom/$\AA^2$\cite{greywall}.  
Above this frozen layer, we place an active layer
of helium atoms that are allowed to move in the simulation.  The sampling
then proceeds as described in Refs. \cite{ceprev}, with the 
modification that effective helium-graphite
interactions are added to the effective action.  We use a starting temperature
of 40 K. 

The approximation that the first-layer atoms can be frozen is 
made in order to concentrate on the second-layer atoms.  Available
computer resources and time constraints make calculations with more
than 50 active particles impractical.  By freezing the first layer,
we can perform calculations with a reasonable number of second-layer
atoms and thus minimize finite-size effects and increase the number
of density values that can be studied.  The trade-off is that we 
ignore the response of the first
layer to the growth of the second.  This is known to 
lower the energy of a layer of helium adsorbed onto solid 
hydrogen\cite{wagner92}.  However, experimental results indicate
that freezing the first layer of helium on a graphite substrate
is a reasonable approximation for the temperatures and densities
of our simulation.  First, the first layer has
a Debye temperature that is greater than 
50 K, and can be treated as a 2D Debye solid for temperatures
as high as 3 K\cite{dash83}.  The temperatures in our 
simulation are as low as 200 mK and never exceed 2.22 K, 
so the first layer is relatively stiff for the conditions in our simulation.
Second, although the first layer is known to be compressed by the growing
second layer, this is most important 
at low second layer densities 
\cite{bretz78}.  The densities of Ref. \cite{bretz78} are below 
the range of our simulation.

Density regions with phase coexistence at zero temperature
can be identified by applying 
the Maxwell construction to the total
ground state energy.  A coexistence region in the thermodynamic
limit will have a total ground state energy that is the weighted average
of the two constituent phases' energy values.  
In Monte Carlo simulations, the energy of the system will lie above the
coexistence line, either because the system remains in an unphysical
homogeneous phase or because creating the phase boundary has a finite 
energy cost.
We can thus identify a coexistence region
as the maximum range of densities in which all the intermediate
energy values lie on or above a line connecting the endpoint values.  

\begin{figure}[htp]
\epsfxsize=\figwidth\centerline{\epsffile{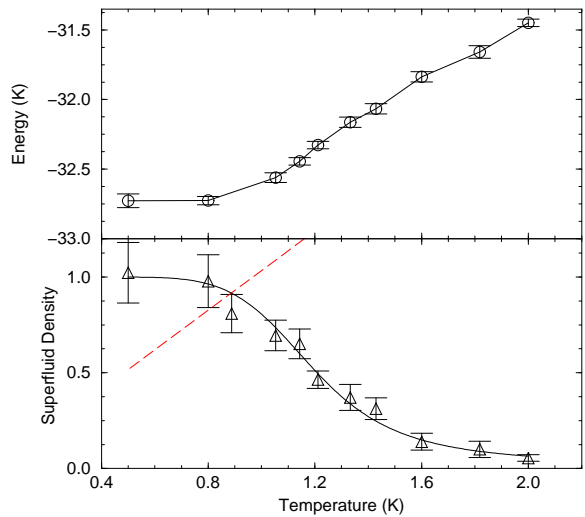}}
\caption{
The temperature dependence of (a) the energy per particle and (b) the
superfluid density in the liquid phase.
}
\label{sf_and_engd178}
\end{figure}

The ground state energy is not directly accessible using PIMC.  We
instead use a limiting process to identify temperatures at which
the system is effectively in the ground state.  
All energy calculations used 
to identify phase regions
in the Maxwell construction 
were performed at 200 mK.  
We then verified that these were
ground state values by recalculating selected values at 400 mK.
In all cases, the values at the two temperatures were within error
bars, indicating that they had converged to 
their zero-temperature values.
See for example Fig. \ref{sf_and_engd178}(a).

\begin{figure}[htp]
\epsfxsize=\figwidth\centerline{\epsffile{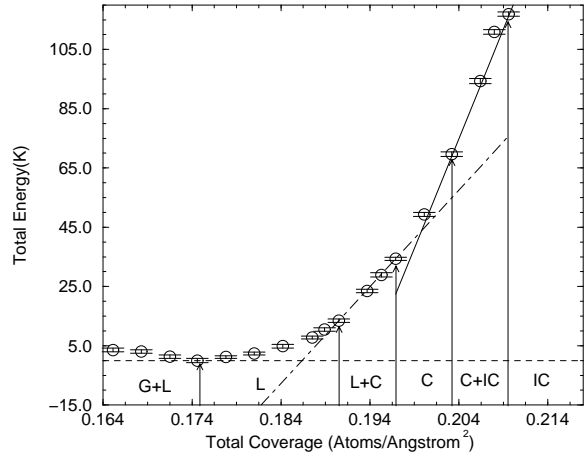}}
\caption{
The total energy found using a $24.12\AA \times 26.11\AA$ simulation
cell with $N_{act}=24,\ldots,\ 52$.
}
\label{bigeng}
\end{figure}

Figure \ref{bigeng} shows the results for the density scans. 
Bounding densities for the L, C, and IC phases
and their coexistence regions at zero temperature
are indicated by vertical arrows.
For clarity, we have subtracted $N_{act}e_{min}$ from the energy values, 
where $N_{act}$ is the number of active particles and
e$_{min}=-32.746 \pm 0.024$ K, the minimum energy per particle.  
All coexistence regions are identified
using the total, not shifted, energy values.
The procedure for identifying these regions is discussed in detail below.

The low density region of the second layer is known experimentally
to be in the G-L phase.
To identify this phase in our simulation, we assume 
that the gas phase at zero temperature has zero density and 
thus zero total energy. 
A coexistence line can then be drawn between 
0.1270 atom/$\AA^{2}$ and 
the density
with the minimum energy per particle, which occurs between 
0.174 and 0.178 atom/$\AA^{2}$.  
This is the dashed line in
Fig. \ref{bigeng}.  The best $\chi^2$ 
parabolic fit around the minimum gives
$\rho_0=0.1750(6)$ atom/$\AA^2$ for the density of minimum energy.  
The number in parenthesis is the error in the last digit.  
We identify the uniform phase region above $\rho_0$ as the L phase
because configurations generated by PIMC 
have no spatial ordering, and the system is 
superfluid at low temperatures, as will be shown
below.  Finite-size
effects on $\rho_0$ are small:  a fit using results from a
significantly smaller cell (approximately one-third the size)
gave $\rho_0=0.1752(6)$, which is the same value within error bars.  
All energy values for the
densities between 0.1270 atom/$\AA^2$ and $\rho_0$ lie above 
the coexistence line, so the system is  
in G-L coexistence for this density range.  If the density of the gas 
phase at zero temperature is not zero then this approach gives a lower 
bound to the end of G-L coexistence.  

The density $\rho_0$ can be compared to experiment.
For $T\leq 0.2$ K the second-layer heat
capacity measurements in Ref. \cite{greywall} show a probable G-L
region roughly between 0.13 and 0.16 atom/$\AA^2$.  Within the
resolution available from the data, this phase may terminate anywhere 
from 0.1600 atom/$\AA^2$ up to, but not including, 0.1700 
atom/$\AA^2$ total coverage.
Since the first-layer coverage 
in the experiment is between
0.120 and 0.127 for these densities, G-L coexistence terminates 
at second-layer coverages anywhere
from 0.033 to 0.050 atom/$\AA^2$.  For comparison, the G-L phase
terminates at the second
layer coverage 0.0480(6) atom/$\AA^2$ in our simulation.  
In the TO measurements, superfluidity is first observed
at 0.174 atom/$\AA^2$, indicating that the superfluid signal in the
experiment becomes significant when the second layer is 
uniformly covered by the superfluid.

Turning now to the highest second-layer densities, we identify another
unstable region, the C-IC phase in Fig. \ref{bigeng}, between 0.2032 and
0.2096 atom/$\AA^2$.  The coexistence line is the 
straight, solid line in the figure that intersects the data
at these two densities.
The intermediate energy values lie on or above this line, so the region
has coexisting phases.  
This coexistence is not a product of 
finite-size effects, since we were able to identify
the same region in
a much smaller simulation cell.
Phase coexistence in fact becomes
clearer in the larger system because we can examine
more density values in the unstable region.  The range 
we find is in good agreement with the coexistence
region 0.2030 to 0.2080 atom/$\AA^2$ that can be determined from
the heat capacity peaks of Ref. \cite{greywall}.

\begin{figure}[htp]
\centering\epsfxsize=\figwidth\centerline{\epsffile{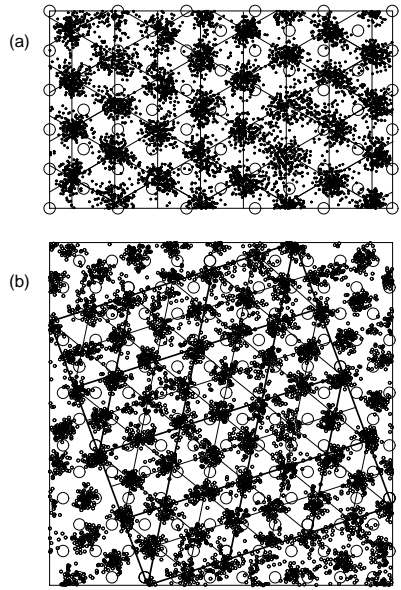}}
\caption{
Snapshots of (a) the incommensurate solid,  
and (b) the $\sqrt{7} \times \sqrt{7}$ commensurate solid.
}
\label{snapshot}
\end{figure}

The higher density phase of this
coexistence region is known experimentally to be an IC solid,
and it is conjectured that the lower density phase is 
a $\sqrt 7 \times \sqrt 7$ C solid.  We can identify these
phases by using simulation cells designed to exactly
accommodate both the first- and second-layer solids.
Figure \ref{snapshot} depicts instantaneous configurations of these
two phases produced by the simulation.  The large circles 
represent first-layer atom
positions, and the small circles show second-layer atom positions
for the configuration.  The solid lines are drawn to emphasize
the triangular structure of both solids.
Figure \ref{snapshot}(a) is an IC phase found  
at 0.2083 atom/$\AA^2$ and 0.2 K.
This phase is incommensurate because no supercell with dimensions
less than the minimum simulation box dimension can be drawn that
has both first- and second-layer atoms periodically repeated.  
Figure \ref{snapshot}(b) depicts an instantaneous configuration of 
the $\sqrt 7 \times \sqrt 7$ C phase at 0.1996 atom/$\AA^2$ and
0.5 K. 
Superlattice unit cells are
indicated by the heavily shaded lines.  Positions of both first and second
layer atoms show a periodic repetition in each superlattice cell.  

The presence of the C phase requires 
an L-C coexistence region between it and the liquid.
The dash-dotted line of Fig. \ref{bigeng} is the L-C coexistence line
found using the Maxwell construction.
Its endpoints 
are 0.1905 and 0.1969 atom/$\AA^2$.  The intermediate energy values lie
on the coexistence line within error bars.
The L-C range is in reasonable agreement 
with the coexistence 
range 0.1871 to 0.1970 atom/$\AA^2$ determined from
heat capacity measurements\cite{greywall}. 
TO measurements also indicate that the L-C
region begins at about 0.187 atom/$\AA^2$\cite{reppy}.

Having identified the L, C, and IC 
phases and their coexistence regions,
we now examine some properties of each phase.
The temperature dependence of the energy and superfluid density
at a sample liquid density, 0.1778 atom/$\AA^{2}$, is given 
in Fig. \ref{sf_and_engd178}.  The values were
calculated using a $15.08\AA \times 15.67\AA$ cell 
with twelve active particles.
The superfluid density is relative to the second-layer density.
The solid curve in Fig. \ref{sf_and_engd178}(b) is the best $\chi^2$ fit to
the solution to the Kosterlitz-Thouless (KT) recursion 
relations \cite{nelson77} integrated to the size of the 
system.  From the intersection of the KT line (dashed
line in the figure) with the fit, we estimate the 
transition temperature to be $T_c \approx 0.88 K$. 

The TO measurements of superfluidity in the second layer possess
unusual features.  No superfluidity can be detected until 
400 mK, and
the superfluid signal never approaches an asymptotic value.
These features can be attributed to both phase coexistence and 
imperfections in the graphite substrate\cite{reppy}.  Our results
support the conclusion of Ref. \cite{reppy} that
if TO measurements can be repeated using a more
uniform graphite substrate, then behavior more typical of a 
2D superfluid, which we find, will be observed.

\begin{figure}[htp]
\epsfxsize=\figwidth\centerline{\epsffile{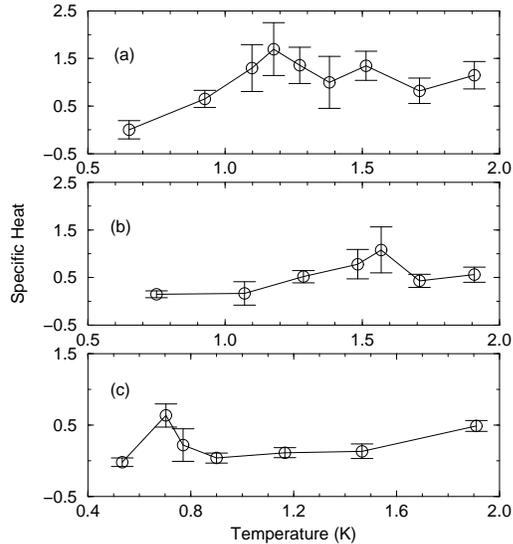}}
\caption{
The specific heats for the (a) liquid, (b) commensurate solid,
and (c) the incommensurate solid.  
}
\label{cvall}
\end{figure}

The specific heat can be obtained by 
differencing the energy with respect to temperature.  Figure \ref{cvall}
shows the results for the L, C, and IC phases.  These can be compared 
to the heat capacity measurements
of Ref. \cite{greywall}.
In the liquid phase, Fig. \ref{cvall}(a), the  
specific heat has a maximum at 
$T=1.18$ K, in fair agreement with the peak at 1 K in the
heat capacity at the same 
coverage.  The results for the $\sqrt{7} \times \sqrt{7}$ solid, 
Fig. \ref{cvall}(b), 
show a peak at 1.57 K, in close agreement with the heat capacity measurements
at a similar coverage.
This provides some additional evidence that
the $\sqrt 7 \times \sqrt 7$ C phase occurs in the experiment.
Figure \ref{cvall}(c) shows the results for the IC solid.
We obtain a peak at 0.70 K.  The peak is at 1 K in the experiment.

This work was supported in part by the National Aeronautics and Space
Administration under Grant No. NAG3-1841.
Some of the calculations were performed using the 
facilities of the Supercomputer Computations Research Institute and
the National High Magnetic Field Laboratory at the Florida State University.


\end{document}